\documentclass[Showpacs,preprintnumbers,amsmath,twocolumn]{revtex4}
\usepackage{graphicx}
\usepackage{dcolumn}
\usepackage{bm}
\usepackage{amssymb}
\usepackage{amsmath}
\usepackage{epsf}
\usepackage{subfigure}
\usepackage{epstopdf}%
\usepackage{amsfonts}

\def\tp{{t^\prime}}

\def\ip{{i^\prime}}
\def\jp{{j^\prime}}
\def\kp{{k^\prime}}
\def\lp{{l^\prime}}
\def\mp{{m^\prime}}
\def\np{{n^\prime}}

\def\cg{{\cal G}}

\def\i{\mbox{i}}
\def\cp{{\cal P}}

\begin{document}

\title{Many-body Green's function approach to attosecond nonlinear X-ray spectroscopy}
\author{Upendra Harbola$^1$ and Shaul Mukamel$^2$}
\affiliation{$^1$Department of Chemistry and Biochemistry, University of California, San Diego,
California 92093-0340, United States\\
$^2$Department of Chemistry, University of California,
Irvine, California 92697-2025, United States}
\date{\today}

\begin{abstract}
Closed expressions are derived for resonant multidimensional X-ray spectroscopy using the quasiparticle nonlinear exciton representation of optical response. This formalism is applied to predict coherent four wave mixing signals which probe single and two core-hole states. Nonlinear X-ray signals are compactly expressed in terms of one- and two- particle Green's functions which can be
obtained from the solution of Hedin-like equations at the $GW$ level.
\end{abstract}

\maketitle

\section{introduction}

With the advent of femtosecond to attosecond X-ray sources \cite{CavalieriPRL2005,seriviceScience2002,schoenleinScience2000,spielmannScience1997,changPRL1997,Murnane2008}, time resolved resonant core level X-ray spectroscopy \cite{kotani} has become a promising tool for studying electron and nuclear motions in real space and real time. Resonant techniques provide detailed information beyond the charge density derived from time-resolved diffraction. Nonlinear X-ray spectroscopy can monitor the dynamics such as making and breaking of chemical bonds at the molecular level.

In a recent work \cite{up-shaulPRB09} we derived closed expressions for attosecond coherent stimulated x-ray Raman signals (CXRS).
It was shown that CXRS can be expressed in terms of one-particle Green's functions which depend parametrically
on the core-hole. In this paper we extend this approach to study nonlinear x-ray spectra by combining the many-body Green's function formulation with the 
exciton representation of valence electronic excitations.
The exciton representation has been particularly useful for computing the optical valence excitations of molecular aggregates
\cite{amerogen,mukamelChemPhys,Darius-Review}. Hedin's Green's function approach \cite{hedin} provides an exact many-body formulation which avoids the computation of many-body wavefunctions. In most applications, Green's function theory requires a self-consistent computation which can become quite tedious and demanding as the system size grows. Various approximations are therefore invoked in practical applications. The $GW$ approximation \cite{hedin,gunnarsson} constitutes a good compromise of accuracy and cost, and various variants have often been used in many applications of the Green's function approach. By combining the exciton modelling of nonlinear response with many-body formulation, we can reduce the complexity of the Green's function approach. 
We derive a Bethe-Salpeter (BS) type equation for two electron-hole pairs. In the standard BS equation\cite{gunnarsson}, the scattering matrix which represents the interaction between an electron and a hole, is expressed as a derivative of the one-particle self-energy with respect to the one-particle Green's function which is then  calculated using the $GW$ approximation \cite{rohlfingPRB2000,strinati}. A further simplification is introduced by taking the screened-coulomb interaction to be independent of the Green's function\cite{strinati}. Despite the lack of a rigorous justification, this approach works well for a broad range of systems including isolated localized atoms and molecules\cite{rohlfingPRB2000}, bulk semiconductors and insulators\cite{albrechtPRL1998,RohlfingPRL81}, conjugated polymers\cite{RohlfingPRL82} and infinite periodic crystals \cite{rohlfingPRB2000,strinati}. It has been difficult to quantify the influence of the variation of screened potential with respect to Green's function on the optical spectra because it involves an expensive self-consistent calculation of the derivative of the screened potential. Here we employ the $GW$ approximation to derive an algebraic expression for the electron-hole scattering matrix which only depends on the one-particle Green's function. The variation of screened-coulomb potential is included approximately. This yields a simplified practical expression for the scattering-matrix which requires comparable efforts to computing the one-particle Green's function at the $GW$ level. We extend this result and combine it with the nonlinear exciton equation (NEE) approach \cite{darius-ChemPhys2005,mukamelPRB75}to compute nonlinear X-ray signals.

The paper is organized as follows. In the next section, we introduce the one- and two-particle Green's functions parametrized in terms of  core-holes. By defining a projection operator in the one-excitation space, we derive a closed expression for the nonlinear
signal in the direction $k_{III}={\bf k}_1+{\bf k}_2-{\bf k}_3$ in terms of parametrized Green's functions. In Sec. (\ref{green}), we present a self-consistent scheme for
 computing these Green's functions. We conclude in Sec. (\ref{conclusion}).

\section{Green's function expression for the nonlinear response}

We divide the electronic system of the molecule into a core and the valence parts. The system is described by the
Mahan-Nozieres-De Dominics (MND) deep core Hamiltonian \cite{RouletPhysRev1969,NozierePhysRev1969,MahanPRB1982}
\begin{eqnarray}
\label{1}
\hat{H}= \hat{H}_0 + \hat{H}_C + \hat{H}_I
\end{eqnarray}
where $\hat{H}_0$ is the free electron part,
\begin{eqnarray}
\label{2}
\hat{H}_0 = \sum_{i}\epsilon_i \hat{c}_i^\dag \hat{c}_i +  \sum_{n}\epsilon_n \hat{c}_n^\dag \hat{c}_n.
\end{eqnarray}
The indices $i,j,k,l$  denote the valence orbitals whereas $m,n$ are the core orbitals. $c_i^\dag$ $(c_i)$ are 
Fock space Fermi creation (annihilation) operators. The coulomb interaction is
\begin{eqnarray}
\label{3}
\hat{H}_C&=& \frac{1}{2} \sum_{ij\ip\jp}V_{ij\ip\jp}\hat{c}^\dag_i\hat{c}_j^\dag\hat{c}_\ip\hat{c}_\jp\nonumber\\
&+&\frac{1}{2} \sum_{mn\mp\np}V_{mn\mp\np}\hat{c}^\dag_m\hat{c}_n^\dag\hat{c}_\mp\hat{c}_\np\nonumber\\
&-& \sum_{imnj}W_{imnj}\hat{c}^\dag_i\hat{c}_m\hat{c}_n^\dag\hat{c}_j.
\end{eqnarray}
The three terms represent respectively valence-valence, core-core and valence-core interactions.
The dipole interaction with the X-ray pulse,
\begin{eqnarray}
\label{4}
\hat{H}_I = -\sum_{im} \left(\frac{}{}E^+(t)\mu^*_{im}\hat{c}_i^\dag\hat{c}_m + E^-(t)\mu_{im}\hat{c}_m^\dag \hat{c}_i\frac{}{}\right)
\end{eqnarray}
where the field is treated as a scalar for simplicity. We assume that the core-holes created by exciting core-electrons to the valence orbitals have an infinite mass. Their motion can thus be ignored and they enter the calculation as fixed parameters. Only valence electron dynamics needs to be considered. This is a reasonable approximation for X-ray spectroscopy where a hole is created in one of the core orbitals which is tightly bound to the atom nucleus compared to the valance electrons. Core migration considerably slows down for deeper core levels\cite{cederbaum}.

\begin{figure}[h]
 \centering
\rotatebox{0}{\scalebox{.45}{\includegraphics{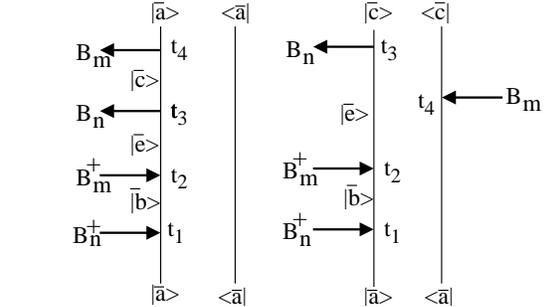}}}\newline
\caption{The nonlinear signal in the direction ${\bf k}_{III}={\bf k}_1+{\bf k}_2-{\bf k}_3$ is depicted using 
the double sided Feynman diagrams, describing the evolution of the density matrix. The time increases from bottom to top.
$|\bar{a}\rangle$ is the ground state of the molecular system. $|\bar{b}\rangle$ and $|\bar{c}\rangle$ are the singly excited states
while $|\bar{e}\rangle$ is the doubly excited many-body state.}
\label{figure1}
\end{figure}

Using the exciton quasiparticle representation, the coherent nonlinear response functions can be
expressed in terms of the exciton scattering-matrix which represents the interaction between two 
excitons created by interactions with X-ray pulses at different sites \cite{shaulPRB2005,Darius-Review}.
For clarity we focus on the double quantum coherence signal generated in the direction ${\bf k}_{III}={\bf k}_1+{\bf k}_2-{\bf k}_3$ \cite{Darius-Review}. 
Other signals are discussed in Appendix \ref{k12}.
Two temporally well separated incoming X-ray pulses interact with the molecule 
and excite two core-electrons at $m$ and $n$ ($m\neq n$) to the
valence orbitals. The ${\bf k}_{III}$ process is depicted by double-sided Fynmann diagram
in Fig. (\ref{figure1})  \cite{shaul-book}. 
%%%%%%%%%%%%%%%%%%%%%%%%%%%%%%%%%%%%%%%%%%%%%%%%%%

%%%%%%%%%%%%%%%%%%%%%%%%%%%%%%%%%%%%%%%%%%%%%%%%%%
The corresponding correlation function expression is given by
\begin{widetext}
\begin{eqnarray}
\label{kIII}
&& S_{III}(t_4,t_3,t_2,t_1) =\theta(t_4-t_3)\theta(t_3-t_2)\theta(t_2-t_1)\nonumber\\
&\times& \sum_{m\neq n}\left[\frac{}{}\left\langle B_m(t_4)B_n(t_3)B^\dag_m(t_2)B^\dag_n(t_1)\right\rangle + \left\langle B_n(t_4)B_m(t_3)B^\dag_m(t_2)B^\dag_n(t_1)\right\rangle + t_4\Leftrightarrow t_3\frac{}{}\right].
\end{eqnarray}
\end{widetext}
The core exciton operators $B$ and $B^\dag$ are defined as
\begin{eqnarray}
\label{bmn}
B_m &=& \sum_{i}\mu_{im}c_ic_m^\dag\nonumber\\
B_m^\dag &=& \sum_{i}\mu_{mi}c_mc_i^\dag.
\end{eqnarray}
The time evolution is in the Heisenberg representation,
\begin{eqnarray}
B(t) = \mbox{e}^{iH^\prime t}B \mbox{e}^{-iH^\prime t}
\end{eqnarray}
where $H^\prime=H_0+H_C$. $t_4\Leftrightarrow t_3$ inside the brackets in Eq. (\ref{kIII}) represents two terms obtained by an interchange of $t_3$ and $t_4$ in the first correlation functions and the expectation value in Eq. (\ref{kIII}) is with respect to the $N$-electron ground state of the full (valence+core) system. 
For clarity we assume that the  x-ray wavelength is short compared to the separation of sites $n$ and $m$ so that the phase variation of the 
field across the molecule may be neglected. This condition may not hold for hard x-rays. Equation (\ref{kIII}) then need to be corrected  by adding phase 
factors  $exp(i{\bf k}.({\bf r}_n- {\bf r}_m)$. The more general expressions are given in the appendix of \cite{pub619}.

A formally straightforward way to compute the signal $S$ is by expanding it in many-body states and their eigenvalues \cite{mukamelPRB75}. However computing the many-body states can become very demanding while efficient electronic structure codes  exist for valence excitations, these are much less developed for core-excitations. Here we use an alternative approach and express $S_{III}$ in terms of the one and two-particle Green's functions. An algorithm is presented for their computation which avoids the explicit computation of the many-body states.

We are interested in the dynamics of two valence electrons in presence of the Coulomb potential generated by core-holes. The core-holes are treated as fixed parameters.

We next introduce the one- and two-electron Green's functions\cite{mills,gdmahan},
\begin{eqnarray}
\label{1-2-gf}
G^{(m)}_{ij}(t,\tp)&=& -i\langle T c_i(t)c_j^\dag(\tp)\rangle_m\nonumber\\
G^{(m,n)}_{ijkl}(t,\tp, t_1,\tp_1)
&=& (-i)^2 \langle T c_i(t)c_j(\tp)c_k^\dag(t_1)c_l^\dag(\tp_1)\rangle_{m,n}\nonumber\\
\end{eqnarray}
where $\langle \cdot\rangle_m$ ($\langle \cdot\rangle_{m,n}$) denote the trace over N-electron ground state in the Coulomb potential of one (two) hole(s) at $m$($m,n$). $T$ is the time ordering operator; when acting on a product of time dependent operators, it 
rearranges them in increasing order of time from right to the left. The superscripts $(m)$ and $(m,n)$ represent parametric dependence on a single core-hole located at $m$ and two core-holes at $m$ and $n$, respectively.
$G_{ij}^{(m)}$ can be expressed as,
\begin{eqnarray}
\label{n-1}
 G_{ij}^{(m)}(t,\tp) &=& -i\theta(t-\tp)\langle a|c_i(t)c_j^\dag(\tp)|a\rangle_m \\
&+&i\theta(\tp-t)\langle a|c_j^\dag(\tp)c_i(t)|a\rangle_m\nonumber\\
&=& -i\theta(t-\tp)\mbox{e}^{iE_a(t-\tp)}\langle a|c_iU(t-\tp)c_j^\dag|a\rangle_m \nonumber\\
&+&i\theta(\tp-t)\mbox{e}^{-iE_a(t-\tp)}\langle a|c_j^\dag U(\tp-t)c_i|a\rangle_m\nonumber
\end{eqnarray}
where $|a\rangle$ is the valence many-body state with energy $E_a$. The time evolution of operators in
(\ref{n-1}) is given by a parametrized Hamiltonian, $U(t)=e^{-iH(m)t}$. $H(m)$ is obtained from (\ref{1})-(\ref{3})
by tracing over core degrees of freedom after replacing $W_{imnj}=\delta_{mn}W_{ij}^{(m)}$ and $V_{mn\mp\np}=\delta_{m\np}\delta_{n\mp}V^{(m,n)}$ where $V^{(m,n)}$
is a parameter and $W_{ij}^{(m)}$ depends on $m$ parameterically. 
Note that $V^{(m,n)}=0$ for $m=n$.
Since the core excitations are stationary, the core operators are assumed
to be time independent. The parametrized Hamiltonian thus obtained is hermitian. In writing the second
equality sign we have further ignored the dependence of $E_a$ on the core-hole at $m$. Thus all dependence to core-hole is
through the time evolution inside the trace, as marked by a subscript $m$.

Similarly, by parametrizing the core-hole, we can express two-particle (two-time) Green's function as
\begin{eqnarray}
 \label{nn-1}
&&G^{(m,n)}_{ijkl}(t,t,\tp,\tp)=\\
 &&(-i)^2\theta(t-\tp) \mbox{e}^{E_a(t-\tp)}\langle a|c_ic_jU(t-\tp)c_k^\dag c_l^\dag|a\rangle_{m,n}\nonumber\\
&+& (-i)^2\theta(\tp-t) \mbox{e}^{E_a(\tp-t)}\langle a|c_k^\dag c_l^\dag U(t-\tp)c_ic_j|a\rangle_{m,n}\nonumber
\end{eqnarray}
where $m\neq n$.

In order to express the signal in terms of the product of one- and two-electrons Green's functions, we need to reduce our many-body space with $(N+1)$ valence electrons. 
This is done by introducing the following projection operator in the $(N+1)$-valence/$1$-core-hole space
\begin{eqnarray}
\label{projection}
\cp_m &=& \sum_i c_i^\dag c_m |\bar{a}\rangle\langle \bar{a}| c_m^\dag c_i
\end{eqnarray}
where $|\bar{a}\rangle$ is a many-body state of the full (valence+core) system.
This projection operator spans a subspace of the many-body space of $N+1$-valence electrons ($i.$$e.$ CI singles.
The full space includes all possible multiples of excitons).
This projection allows us to obtain a closed expression for the signal in terms of the one- and two-electron Green's functions, which can then 
be computed using many-body techniques.

We next insert the projection operators (\ref{projection}) into the expectation values in Eq. (\ref{kIII}).
This is our key approximation which allows us to express the signal in terms of the product of Green's functions.
Using Eq. (\ref{bmn}) in (\ref{kIII}) and inserting the projection operators we obtain
\begin{widetext}
\begin{eqnarray}
\label{n-newk3}
&&S_{III}(t_4,t_3,t_2,t_1)= \sum_{m\neq n}\sum_{ijkl} \mu_{im}\mu_{mk}\mu_{jn}\mu_{nl} \theta(t_4-t_3)\theta(t_3-t_2)\theta(t_2-t_1) \nonumber\\
&\times&\mbox{e}^{iE_a(t_4-t_1)}\left[ \langle \bar{a}|c_ic_m^\dag U(t_4-t_3){\cal P}_m c_jc_n^\dag U(t_3-t_2) c_m c_k^\dag {\cal P}_n U(t_2-t_1) c_n c_l^\dag|\bar{a}\rangle\right.\nonumber\\
&+&\left. \langle \bar{a}|c_jc_n^\dag U(t_4-t_3){\cal P}_n c_ic_m^\dag U(t_3-t_2) c_m c_k^\dag {\cal P}_n U(t_2-t_1) c_nc_l^\dag|\bar{a}\rangle + t_4 \Leftrightarrow t_3\right].
\end{eqnarray}
\end{widetext}
The exact expression for the signal may be obtained by removing projection operators inside the brackets in Eq. (\ref{n-newk3}).

Using (\ref{projection}), each term on the r.h.s. of Eq. (\ref{n-newk3}) factorizes into a product of three
correlation functions. By parametrizing the core variables, each correlation function can be approximated as
\begin{eqnarray}
 \label{nn-2}
\langle \bar{a}|c_m^\dag c_i U(t_4-t_3)c_j^\dag c_m|\bar{a}\rangle \approx
\langle a|c_i U(t_4-t_3)c_j^\dag|a\rangle_m
\end{eqnarray}
where $U(t_4-t_3)$ on the r.h.s depends on core variables parametrically.

Making use of Eqs. (\ref{n-1}), (\ref{nn-1}) and (\ref{nn-2}), Eq. (\ref{n-newk3}) becomes
%\begin{widetext}
\begin{eqnarray}
\label{finalkIII}
&&S_{III}(t_4,t_3,t_2,t_1)=\theta(t_4-t_3)\theta(t_3-t_2)\theta(t_2-t_1)\nonumber\\
&&\sum_{m\neq n}\sum_{ijkl}\sum_{\ip\jp}\mu_{im}\mu_{mk}\mu_{jn}\mu_{nl}
G_{\ip l}^{(n)}(t_2,t_1)\nonumber\\
&\times& \left[\frac{}{}
 G_{i\jp}^{(m)}(t_4,t_3)\cg_{\jp jk\ip}^{(m,n)}(t_3,t_2)
+ G_{j\jp}^{(n)}(t_4,t_3)\cg_{\jp ik \ip}^{(m,n)}(t_3,t_2)\right.\nonumber\\
&+&\left.(t_4 \Leftrightarrow t_3)\frac{}{}\right].\nonumber\\
\end{eqnarray}
%\end{widetext}
This is a closed expression for the signal in terms of the one- and two- electron Green's functions.
corresponding expressions for other techniques $S_I$ at ${\bf k}_I=-{\bf k}_1+{\bf k}_2+{\bf k}_3$ and $S_{II}$ at ${\bf k}_{II}={\bf k}_1-{\bf k}_2+{\bf k}_3$ are given in Appendix \ref{k12}.

Note that the Green's functions in Eq. (\ref{finalkIII}) depend only on the ddifferences of their time arguments [see Eq. (\ref{n-1})].
Denoting the time delay between x-ray pulses as $\tau_i=t_{i+1}-t_i$, where $i=1,2,3$, and taking the Fourier transform with respect to delay times, the signal can be expressed in the frequency domain as
\begin{eqnarray}
 \label{new}
S_{III}(\omega_3,\omega_2,\omega_1) &=& \int_{-\infty}^{\infty}d\tau_1\int_{-\infty}^{\infty} d\tau_2\int_{-\infty}^{\infty}d\tau_3
\mbox{e}^{i(\omega_1\tau_1+\omega_2\tau_2+\omega_3\tau_3)}\nonumber\\
&&S_{III}(\tau_1+\tau_2+\tau_3,\tau_1+\tau_2,\tau_1,0) ,
\end{eqnarray}
we get
\begin{widetext}
\begin{eqnarray}
\label{finalkIII-freq}
S_{III}(\omega_3,\omega_2,\omega_1) &=& i \sum_{m\neq n}\sum_{ijkl}\sum_{\ip\jp}\mu_{im}\mu_{mk}\mu_{jn}\mu_{nl} \int\int\int \frac{d\omega_1^\prime d\omega_2^\prime d\omega_3^\prime}{(2\pi)^3}
\left(\frac{1}{\omega_3-\omega_3^\prime-i\epsilon}+\frac{1}{\omega_3+\omega_3^\prime-\omega_2^\prime-i\epsilon}\right)\nonumber\\
&\times&\frac{G_{\ip l}^{(n)}(\omega_1^\prime)[G_{i\jp}^{(m)}(\omega_3^\prime)\cg_{\jp jk\ip}^{(m,n)}(\omega_2^\prime)+ G_{j\jp}^{(n)}(\omega_3^\prime)\cg_{\jp ik \ip}^{(m,n)}(\omega_2^\prime)]}
{(\omega_1-\omega_1^\prime-i\epsilon)(\omega_2-\omega_2^\prime-i\epsilon)}
\end{eqnarray}
\end{widetext}
where $\epsilon$ is an infinitesimal number and
\begin{eqnarray}
 \label{GF-freq}
G_{ij}^{(m)}(\omega) = \int_{-\infty}^\infty dt \mbox{e}^{-i\omega t} G_{ij}^{(m)}(t)
\end{eqnarray}
is the Fourier transform of the Green's function.

\section{computing the Green's functions}\label{green}

The one-electron Green's function, which appears in Eq. (\ref{finalkIII}), can be obtained from the
self-consistent solution of Dyson equation,
\begin{eqnarray}
\label{21aa}
G_{ij}^{(m)}(t,\tp) &=& G_{ij}^0(t,\tp)\nonumber\\
&+& G_{i\jp}^0(t,\tp_1)\tilde{\Sigma}^{(m)}_{\jp\ip}(\tp_1,\tp_2)
G_{\ip j}^{(m)}(\tp_2,\tp)
\end{eqnarray}
where $G^0$ is the reference Green's function corresponding to non-interacting electron system and
\begin{eqnarray}
\tilde{\Sigma}^{(m)} = V_C^{(m)}+\Sigma^{(m)}
\end{eqnarray}
where $V_C^{(m)}$ is the Coulomb potential due to the presence of a core-hole at $m$. $\Sigma^{(m)}$ is the time-dependent self-energy which comes from the electron-electron interactions and satisfy closed exact equations 
(\ref{closed-eq-1})-(\ref{closed-eq-4}). A self-consistent solution of the exact equations is highly demanding numerically. 
The $GW$ approximation is generally used in order to simplify these equations by ignoring the vertex corrections and keeping only the first 
term in Eq. (\ref{closed-eq-4}), the self-consistent calculation of the Green's functions and the self-energy is greatly reduced.
We then have $\Lambda_{ijkl}^{(m)}(t,\tp,t_1) = \delta_{ik}\delta_{lj}\delta(t-t_1)\delta(t-\tp)$ and Eqs. (\ref{closed-eq-2}) and (\ref{closed-eq-4}) reduce to
\begin{eqnarray}
\Sigma_{ij}^{(m)}(t,\tp)&=& \i S^{(m)}_{\ip i \jp j}(\tp,t)G^{(m)}_{\ip\jp}(t,\tp)\label{n-sig-gw}\\
S^{(m)}_{ijkl}(t,\tp) &=& \bar{V}_{iklj}\delta(t-\tp)+ \i \bar{V}_{\ip lk\jp}S^{(m)}_{ij\kp\lp}(\tp_1,\tp)
\nonumber\\
&\times& G^{(m)}_{\ip\kp}(t,\tp_1) G^{(m)}_{\jp\lp}(\tp_1,t)\label{nn-s-gw}
\end{eqnarray}
where $\bar{V}_{ijkl}= V_{ijkl}-V_{jikl}$.
Thus for one-electron Green's function we only need to solve Eq.(\ref{21aa}) self-consistently together with (\ref{n-sig-gw}) and (\ref{nn-s-gw}).
This is a standard approximation which has been used extensively in studying optical properties of different kinds of systems ranging from single atoms to semiconductor clusters and periodic crystals. The FEFF code\cite{rehr} computes one-particle Green's functions by solving the $GW$ equations, Eqs. (\ref{21aa})-(\ref{nn-s-gw}).

The two-electron Green's function, Eq. (\ref{1-2-gf}), satisfies the equation (see Appendix \ref{GFun})
\begin{widetext}
\begin{eqnarray}
\label{2-electron}
\cg_{ijkl}^{(m,n)}(t,\tp,t_1,\tp_1) &=& G_{il}^{(m,n)}(t,\tp_1)G_{kj}^{(m,n)}(t_1,\tp)\nonumber\\
&+& G_{i\ip}^{(m,n)}(t,\tp_2)G_{\jp j}^{(m,n)}(\tp_3,\tp)\Xi_{\ip\jp\kp\lp}^{(m,n)}(\tp_2,\tp_3,\tp_4,\tp_5)
\cg^{(m,n)}_{\kp\lp kl}(\tp_4,\tp_5,t_1,\tp_1).
\end{eqnarray}
\end{widetext}
where the exact interaction kernel, $\Xi$, is given by Eq. (\ref{kernel-1}).
Note that one-electron Green's functions entering in (\ref{2-electron}) is calculated in the presence of two core-holes at $m$ and $n$. These satisfy equations similar to Eqs. (\ref{21aa}), (\ref{n-sig-gw}) and (\ref{nn-s-gw}) which can be obtained by simply replacing superscripts $(m)$ or $(n)$ by $(m,n)$ and using $\tilde{\Sigma}^{{(m,n)}}=V_C^{(m)}+V_C^{(n)}+\Sigma^{(m,n)}$.
Recently Feng $et$ $al$ \cite{fangPRB08} have used nonresonant X-ray Raman scattering to study exciton spectroscopy on $h-BN$ near boron $K$-edge by computing the two-particle Green's function.
Within the $GW$ approximation, the interaction kernel for the two-electron Green's function, Eq. (\ref{kernel-1}), reduces to a simpler form,
\begin{eqnarray}
\label{gamma-24}
&&\Xi_{ijkl}^{(m,n)}(t,\tp,t_1,\tp_1)
= \i S^{(m,n)}_{kilj}(\tp,t)\delta(t-t_1)\delta(\tp_1-\tp)\nonumber\\
&-& \bar{V}_{\ip j\jp l}S^{(m,n)}_{\ip i\kp k}(t_1,t)G^{(m,n)}_{\ip\kp}(\tp,t_1)\delta(\tp_1-\tp)\nonumber\\
&-& \bar{V}_{kj\jp\ip}S^{(m,n)}_{\ip il\kp}(\tp_1,t)G^{(m,n)}_{\kp\ip}(\tp_1,\tp)\delta(\tp-t_1).\nonumber\\
\end{eqnarray}
%\end{widetext}
This is our final result for the scattering matrix between two valence electrons in presence of two core-holes.
The first term represents the contribution from screened coulomb interaction and the other two terms are induced by the
change in screening of the electron in valance and hole in the core regions. 
Equations (\ref{2-electron}) and (\ref{gamma-24}) together with Eqs. (\ref{21aa}), (\ref{n-sig-gw}) and (\ref{nn-s-gw}) (after making above changes in superscripts for two core-holes) constitute a closed set of equations which can be solved self-consistently to obtain the two particle Green's function. 
Since both $S^{(m,n)}_{ijkl}$ and $G^{(m,n)}_{ik}$ come from the self-consistent calculation of the Dyson equation in presence of two-core 
holes, $\Xi^{(m,n)}_{ijkl}$ can be readily computed from Eq. (\ref{gamma-24}). The numerical effort involved 
in computing the two-electron scattering-matrix is comparable to computing the one-particle Green's function.

\section{conclusion}\label{conclusion}
We have derived closed expressions for the nonlinear X-ray spectra in terms of the many-body Green's functions. 
We expressed the ${\bf k}_{III}={\bf k}_1+{\bf k}_2-{\bf k}_3$ signal in terms of one- and two-particle Green's functions. 
In the deep core Hamiltonian formulation, the slow core-hole dynamics is ignored as compared to the fast time evolution of the valence electron system, and
core-holes are simply treated as parameters. The key approximation was the introduction of a projection operator inside the multipoint correlation function of exciton operator which allows to express the signal in terms of one and two-particle Green's functions in a simple way. In order to compute these Green's functions, we have generalized the Hedin's equations \cite{hedin} and derived a modified Bethe-Salpeter equation for two-particle Green's function in presence of two-core holes. A simple expression for the scattering matrix for two electrons is derived in terms of one-particle Green's function which includes the effect of the change in screened potential which goes beyond the usual $GW$ approximation where variation in screening potential is generally ignored. The present formulation can be generalized straightforwardly to accomodate inelastic interactions on the X-ray nonlinear signals by including their effects through self-energy for one and two particle Green's functions.

\section*{ACKNOWLEDGMENTS}

The support of the Chemical Sciences, Geosciences and Biosciences Division, Office of Basic Energy Sciences, Office of Science and U.S. Department of Energy is gratefully acknowledged.

\appendix

%%%%%%%%%%%%%%%%%%%%%%%%%%%%%%%%%%%%%%%%%%%%%%%%%%%%%%%%%%%%%%%%%%%%%%%%%%%%%%%%%%%%%%%%%%%%%%%%%%%%%%%%%%%%%%%%%%%%%
%%%%%%%%%%%%%%%%%%%%%%%%%%%%%%%%%%%%%%%%%%%%%%%%%%%%%%%%%%%%%%%%%%%%%%%%%%%%%%%%%%%%%%%%%%%%%%%%%%%%%%%%%%%%%%%%%%%%%

\section{${\bf k}_I$ and ${\bf k}_{II}$ signals}\label{k12}

\begin{figure}[h]
 \centering
\rotatebox{0}{\scalebox{.35}{\includegraphics{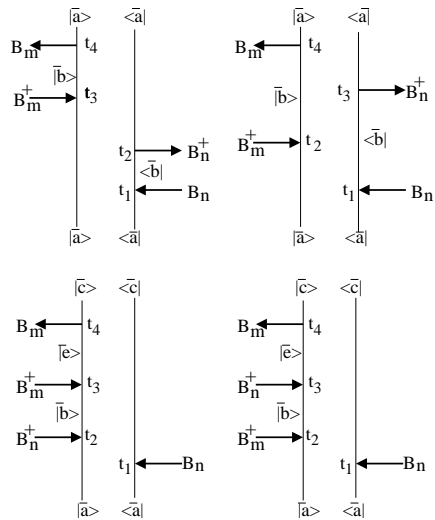}}}\newline
\caption{Double sided Feynman diagram for the nonlinear signal in the direction 
${\bf k}_{I}=-{\bf k}_1+{\bf k}_2+{\bf k}_3$. Time increases from bottom to top.
$|\bar{a}\rangle$ is the ground state of the molecular system. $|\bar{b}\rangle$ and $|\bar{c}\rangle$ are the singly excited states
while $|\bar{e}\rangle$ is the doubly excited many-body state.}
\label{figure2a}
\end{figure}

\begin{figure}[h]
 \centering
\rotatebox{0}{\scalebox{.35}{\includegraphics{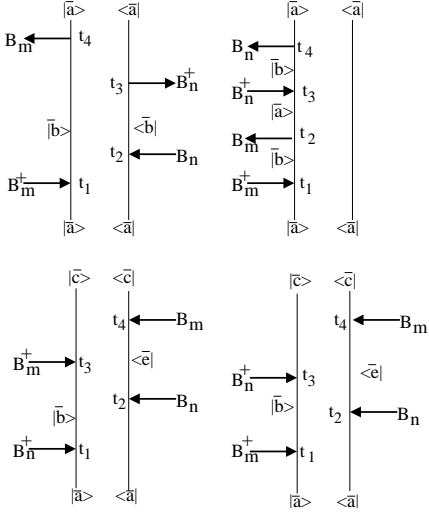}}}\newline
\caption{Double sided Feynman diagrams for nonlinear signal in the direction $K_{II}={\bf k}_1-{\bf k}_2+{\bf k}_3$.}
\label{figure2b}
\end{figure}

The ${\bf k}_I-{\bf k}_1+{\bf k}_2+{\bf k}_3$ and ${\bf k}_{II}={\bf k}_1-{\bf k}_2+{\bf k}_3$ processes represent the 
interaction with two X-ray pulses and involve four Liouville space pathways which
can be expressed in terms of double sided Feynman diagrams \cite{shaulPRB2005} shown in Figs. (\ref{figure2a}) and (\ref{figure2b}). The corresponding expressions in terms of the correlation functions of the exciton variables analogous to (\ref{kIII}) are given by
%%%%%%%%%%%%%%%%%%%%%%%%%%%%%%%%%%%%%%%%%%%%%%%%%%
%%%%%%%%%%%%%%%%%%%%%%%%%%%%%%%%%%%%%%%%%%%%%%%%%%
%%%%%%%%%%%%%%%%%%%%%%%%%%%%%%%%%%%%%%%%%%%%%%%%%%
%%%%%%%%%%%%%%%%%%%%%%%%%%%%%%%%%%%%%%%%%%%%%%%%%%
\begin{eqnarray}
S_I(t_1,t_2,t_3,t_4) &=& \sum_{mn}\theta(t_4-t_3)\theta(t_3-t_2)\theta(t_2-t_1)\nonumber\\
&&\left[\langle B_n(t_1)B_n^\dag(t_2)B_m(t_4)B_m^\dag(t_3)\rangle\right.\nonumber\\
&+&\left.\langle B_n(t_1)B_n^\dag(t_3)B_m(t_4)B_m^\dag(t_2)\rangle\right.\nonumber\\
&+& \left. \langle B_n(t_1)B_m^\dag(t_4)B_m^\dag(t_3)B_n^\dag(t_2)\rangle\right.\nonumber\\
&+&\left.\langle B_n(t_1)B_m(t_4)B_n^\dag(t_3)B_m^\dag(t_2)\rangle
\right]\label{ex-k1}
\end{eqnarray}
and
\begin{eqnarray}
S_{II}(t_1,t_2,t_3,t_4) &=& \sum_{m\neq n}\theta(t_4-t_3)\theta(t_3-t_2)\theta(t_2-t_1)\nonumber\\
&&\left[\langle B_n(t_2)B_n^\dag(t_3)B_m(t_4)B_m^\dag(t_1)\rangle\right.\nonumber\\
&+&\left.\langle B_n(t_4)B_n^\dag(t_3)B_m(t_2)B_m^\dag(t_1)\rangle\right.\nonumber\\
&+& \left. \langle B_n(t_2)B_m(t_4)B_m^\dag(t_3)B_n^\dag(t_1)\rangle\right.\nonumber\\
&+&\left.\langle B_n(t_2)B_m(t_4)B_n^\dag(t_3)B_m^\dag(t_1)\rangle
\right]\label{ex-k2}
\end{eqnarray}
where $t_i, i=1,2,3,4$ is the interaction time of the four X-ray pulses, and operator $B_m$ is defined in Eq. (\ref{bmn}).
Repeating the steps that transformed Eq. (\ref{kIII}) to (\ref{finalkIII}), we can express these signals in terms of the
one and two-particle Green's functions as
\begin{eqnarray}
\label{k1-ex}
&& S_I(t_1,t_2,t_3,t_4) = \sum_{mn}\sum_{a}P(a)\sum_{ijkl}\sum_{\ip\jp} \mu_{im}\mu_{mj}\mu_{kn}\mu_{nl}\nonumber\\
&\times&\theta(t_4-t_3)\theta(t_3-t_2)\theta(t_2-t_1)
\left[\frac{}{}\sum_{c}\{f_{c;l\ip}f^*_{c;i\jp}\mbox{e}^{-iE_{ca}(t_2-t_4)}\right.\nonumber\\
&\times& \left. G_{k\ip}^{(n)\dag}(t_2,t_1)G_{\jp j}^{(m)}(t_4,t_3)+t_2\Leftrightarrow t_3\}
-G_{k\ip}^{(n)\dag}(t_4,t_1) \right.\nonumber\\
&\times&\left. \left(G_{\jp l}^{(n)}(t_3,t_2)\cg_{\ip i\jp j}^{m,n}(t_4,t_3)
+G_{\jp j}^{(m)}(t_3,t_2)\cg_{\ip il\jp}^{m,n}(t_4,t_3) \right)\!\!\!\frac{}{}\right].
\end{eqnarray}
Similarly for ${\bf k}_{II}$ we obtain,
\begin{eqnarray}
\label{k2-ex}
&& S_{II}(t_1,t_2,t_3,t_4) = \sum_{m\neq n}\sum_{a}P(a)\sum_{ijkl}\sum_{\ip\jp} \mu_{im}\mu_{mj}\mu_{kn}\mu_{nl}\nonumber\\
&\times&\theta(t_4-t_3)\theta(t_3-t_2)\theta(t_2-t_1)
\left[\frac{}{}\sum_{c}\{f_{c;l\ip}f^*_{c;i\jp}\mbox{e}^{-iE_{ca}(t_3-t_4)}\right.\nonumber\\
&\times& \left. G_{k\ip}^{(n)\dag}(t_3,t_2)G_{\jp j}^{(m)}(t_4,t_1)+t_2\Leftrightarrow t_4\}
-G_{k\ip}^{(n)\dag}(t_4,t_2) \right.\nonumber\\
&\times&\left. \left(G_{\jp l}^{(n)}(t_3,t_1)\cg_{\ip i\jp j}^{m,n}(t_4,t_3)
+G_{\jp j}^{(m)}(t_3,t_1)\cg_{\ip il\jp}^{m,n}(t_4,t_3) \right)\!\!\!\frac{}{}\right]
\end{eqnarray}
where $f_{c;ij}$ are the expansion coefficients in terms of the configuration interaction 
singles \cite{up-shaulPRB09}.

\section{Derivation of Eqs. (\ref{21aa}) - (\ref{2-electron})}\label{GFun}

We use a Hamiltonian parametrized in terms of the core-holes obtained by tracing out the core-hole degrees of freedom.
\begin{eqnarray}
\label{par1}
H= \sum_{i}\epsilon_i c_i^\dag c_i + \sum_{ijkl}V_{ijkl}c_i^\dag c_j^\dag c_k c_l + \sum_{ij}\phi_{ij}^{(m)} c_i^\dag c_j.
\end{eqnarray}
The first two terms represent the kinetic energy and the electron-electron interaction, $\phi_{ij}^{m}$ in the last term represents the potential
 due to the presence of a core-hole located at $m$th core orbital. It is obtained by approximating the last term in Eq. (\ref{3}) with,
\begin{eqnarray}
\label{par2}
\phi_{ij}^{(m)} \sim -W_{immj} c_i^\dag c_j \langle c_m c_m^\dag \rangle.
\end{eqnarray}
We shall treat the valence electrons in the (average) field of two  core-holes.
We are interested in  the dynamics of one and two electrons created by excitation from the core orbitals.
We thus need to calculate
\begin{eqnarray}
\label{par3}
\chi_{ijkl}(t,\tp,t_1\tp_1) = \langle T c_i(t)c_j^\dag (\tp)c_k(t_1) c_l^\dag(\tp_1)\rangle.
\end{eqnarray}
which for $t,t_1>\tp_1,\tp$ represents the dynamics of two electrons created at times $t_1$ and $\tp$ and destroyed at times $t$ and $\tp_1$. $\chi_{ijkl}$ can be connected to a four-point Green's function $\cg_{ijkl}^{(m,n)}$ as
\begin{eqnarray}
\label{par4}
\cg_{ijkl}(t,\tp,t_1\tp_1) = i\chi_{ijkl}(t,\tp,t_1\tp_1)+iG_{ij}(t,\tp)G_{kl}(\tp_1,t_1).
\end{eqnarray}
In order to obtain a closed equation for $\tilde{G}^{(m)}$ and $\cg_{ijkl}^{(m,n)}$, we add the following source term to the Hamiltonian, Eq. (\ref{par1}).
\begin{eqnarray}
H_G=\sum_i\eta_i c_i + \sum_i\eta^*c_i^\dag
\end{eqnarray}
Here $\eta_i$ and $\eta^*_i$ are
Grassmann variables which satisfy,
\begin{eqnarray}
\label{grassmann}
\frac{\partial}{\partial \eta_i} \frac{\partial}{\partial \eta_j^*}= - \frac{\partial}{\partial \eta_j^*}\frac{\partial}{\partial \eta_i}.
\end{eqnarray}
We further define a generalized one-particle Green's function
\begin{eqnarray}
\label{gen-gf}
\tilde{G}_{ij}(t,\tp) := -\frac{i}{I(\eta,\eta^*)}\langle T c_i(t)c_j^\dag(\tp)\rangle_\eta
\end{eqnarray}
where $\langle\cdot\rangle_\eta$ represents the trace with respect to the density matrix corresponding to the full Hamiltonian including the Grassmann terms and
\begin{eqnarray}
\label{i}
I(\eta,\eta^*) = \left\langle T \mbox{e}^{-i\int d\tau \sum_i(\eta_i(\tau) c_i(\tau)+\eta_i^*(\tau)c_i^\dag(\tau))}\right\rangle_\eta
\end{eqnarray}
The one-particle Green's function $G_{ij}^{(m,n)}$ is obtained simply by setting $\eta=\eta^*=0$ in Eq. (\ref{gen-gf}). Switching over to the interaction picture with respect to the Hamiltonian $H_G$, Eq. (\ref{gen-gf}) can be expressed as,
\begin{eqnarray}
\label{int-gen-gf}
\tilde{G}_{ij}(t,\tp) = -\frac{i}{I(\eta,\eta^*)}\left\langle T c_i(t)c_j^\dag(\tp)I(\eta,\eta^*)\right\rangle
\end{eqnarray}
Taking the derivative with respect to $\eta$ and $\eta^*$, we obtain,
\begin{eqnarray}
\label{111}
\left.\frac{\partial^2\tilde{G}_{ij}(t,\tp)}{\partial\eta^*_l(\tp_1)\partial\eta_k(t_1)}\right|_{\{\eta\}=0} = \cg_{ijkl}(t,\tp,t_1,\tp_1)
\end{eqnarray}
where $\{\eta\}=0$ is the short-hand notation for $\eta=\eta^*=0$.

Thus our strategy is to first compute a closed equation for the Green's function $\tilde{G}_{ij}$ and then using Eq. (\ref{111}), we obtain a closed equation for the four particle Green's function, $\cg_{ijkl}$. 

The equation of motion for $\tilde{G}_{ij}$ is derived by taking time derivative of Eq. (\ref{gen-gf}).
\begin{eqnarray}
\label{eom-g}
i I(\eta,\eta^*)\frac{\partial}{\partial t} \tilde{G}_{ij}(t,\tp) &:=& \delta(t-\tp)\delta_{ij}\nonumber\\
&+& \left\langle T \left(\frac{\partial}{\partial t}c_i(t)\right) c_j^\dag(\tp)\right\rangle_\eta.
\end{eqnarray}
The time evolution of $c_i(t)$ is governed by the total Hamiltonian $H+H_G$ through the Heisenberg equation
\begin{eqnarray}
\label{eom-ci}
&& i\frac{\partial}{\partial t} c_i(t) = \epsilon_i c_i(t) + \phi^{m,n}_{i\jp}c_\jp (t)
+ 2\eta_\ip^*(t)c_i(t)c_\ip^\dag(t)\nonumber\\
&-& 2\eta_\ip(t) c_\ip(t)c_i(t) - \bar{V}_{\ip i\jp\kp} c_\ip^\dag(t)c_\jp(t)c_\kp(t) - \eta^*_i\nonumber\\
\end{eqnarray}
where $\bar{V}_{ijkl} = (V_{ijkl}-V_{jikl})/2$.
Substituting Eq. (\ref{eom-ci}) in (\ref{eom-g}), we obtain,
\begin{widetext}
\begin{eqnarray}
\label{eom-g-1}
&&i I(\eta,\eta^*)\left[\left(\frac{\partial}{\partial t} +i\epsilon_i\right)\delta_{\jp i}
+i\phi_{i\jp}^{m,n}(t)\right]\tilde{G}_{\jp j}(t,\tp)
=\delta_{ij}\delta(t-\tp)+i\eta_i^*\langle c_j^\dag(\tp)\rangle_\eta \nonumber\\
&+& i \bar{V}_{\ip i\jp\kp}\langle T c_\ip^\dag(t)c_\jp(t)c_\kp(t)c_j^\dag(\tp)\rangle_\eta
-2i \eta^*_\ip(t)\langle T c_i(t)c_\ip^\dag(t)c_j^\dag(\tp)\rangle_\eta
+2i \eta_\ip(t)\langle T c_\ip(t)c_i(t)c_j^\dag(\tp)\rangle_\eta.
\end{eqnarray}
\end{widetext}
We next connect the nonlinear terms in Eq. (\ref{eom-g-1}) to the one-particle Green's function. By taking first and second derivatives of $G_{ij}$ with respect to $\eta_i$, we get
%\begin{widetext}
\begin{eqnarray}
\label{derivative}
\frac{\delta \tilde{G}_{\kp j}(t,\tp)}{\delta\eta_\jp(t)} &=& -\langle T c_\jp(t)c_\kp(t)c_j^\dag(\tp)\rangle_\eta\nonumber\\
&+&\langle Tc_\kp(t)c_j^\dag(\tp)\rangle_\eta\langle c_\jp(t)\rangle_\eta\nonumber\\
\frac{\delta^2 \tilde{G}_{\kp j}(t,\tp)}{\delta \eta_\ip^*(t)\delta\eta_\jp(t)} &=&
i\langle T c_\ip^\dag(t)c_\jp(t)c_\kp(t)c_j^\dag(\tp)\rangle_\eta \nonumber\\
&-&i\tilde{G}_{\kp j}(t,\tp)\tilde{G}_{\jp\ip}(t,t^+)\nonumber\\
&+&i\frac{\delta \tilde{G}_{\kp j}(t,\tp)}{\delta\eta_\jp(t)}\langle c_\ip^\dag(t)\rangle_\eta\nonumber\\
&+&i\frac{\delta \tilde{G}_{\kp j}(t,\tp)}{\delta\eta_\ip^*(t)}\langle c_\jp(t)\rangle_\eta.
\end{eqnarray}
%\end{widetext}
Using Eqs. (\ref{derivative}) in (\ref{eom-g-1}) we finally get
\begin{eqnarray}
\label{dyson-1}
\tilde{G}_{i\kp}^{0}(t,t_1)\tilde{G}_{\kp j}(t_1,\tp)&=&\delta_{ij}\delta(t-\tp)\nonumber\\
&+&\tilde{\Sigma}_{i\kp}(t,t_1)\tilde{G}_{\kp j}(t_1,\tp)
\end{eqnarray}
where
\begin{eqnarray}
\label{g0}
{\tilde{G}^{0-1}_{ij}}(t,\tp)= \left[\left(i\frac{\partial}{\partial t}
-\epsilon_i\right)\delta_{ij}+v_{ij}^{(m,n)}(t)\right]\delta(t-\tp)
\end{eqnarray}
with
%\begin{widetext}
\begin{eqnarray}
\label{hartree}
v_{ij}^{(m,n)}(t)&=&\left[ 2i\eta_\ip^*(t)\left(\langle c_\ip^\dag(t)\rangle_\eta+\frac{\delta}{\delta\eta_\ip^*(t)}\right)\right.\nonumber\\
&+&\left. 2i\eta_\ip(t)\left(\langle c_\ip(t)\rangle_\eta+\frac{\delta}{\delta\eta_\ip(t)}\right)\right]\delta_{ij}\nonumber\\ &+&\phi_{ij}^{(m,n)}(t)-i\bar{V}_{\ip i\jp\kp}\tilde{G}_{\jp\ip}(t,t^+).
\end{eqnarray}
%\end{widetext}
Note that for $\{\eta\}=0$ the first two terms in (\ref{hartree}) vanish and $v_{ij}^{mn}$ is the sum of Hartree and Coulomb potential due to two core-holes.
The function $\tilde{\Sigma}_{ij}$ in (\ref{dyson-1}) is
%\begin{widetext}
\begin{eqnarray}
\label{sigma}
&&\tilde{\Sigma}_{i\kp}(t,t_1)\tilde{G}_{\kp j}(t_1,\tp) = -i\eta_i^*(t)\langle c_j^\dag(\tp)\rangle \nonumber\\
&+&\bar{V}_{\ip i\jp\kp}\frac{\delta^2\tilde{G}_{\kp j}(t,\tp)}{\delta\eta_\ip^*(t)\delta\eta_\jp(t)}
-i \bar{V}_{\ip i\jp\kp} \frac{\delta \tilde{G}_{\kp j}(t,\tp)}{\delta\eta_\jp(t)}\langle c_\ip^\dag(t)\rangle_\eta\nonumber\\
&-&i \bar{V}_{\ip i\jp\kp} \frac{\delta \tilde{G}_{\kp j}(t,\tp)}{\delta\eta_\ip^*(t)}\langle c_\jp(t)\rangle_\eta.
\end{eqnarray}
%\end{widetext}
When $\{\eta\}=0$, all terms, except the second term which is given by Eq. (\ref{111}), vanish and $\tilde{\Sigma}$ reduces to the self-energy $\Sigma$ due to electron-electron interaction.

In order to get a closed set of equations for the self-energy $\Sigma$, we make use of the
identity,
\begin{eqnarray}
\label{identity}
\tilde{G}_{i\jp}^{-1}(t,t_1)l\tilde{G}_{\jp j}(t_1,\tp)=\delta_{ij}\delta(t-\tp).
\end{eqnarray}
Differentiating Eq.(\ref{identity}) once with respect to $\eta_\jp(t)$, we can write
\begin{eqnarray}
\label{diff-1}
\tilde{G}_{i\kp}^{-1}(t,t_1)\frac{\delta \tilde{G}_{\kp j}(t,\tp)}{\delta\eta_\jp(t)}
=- \frac{\delta \tilde{G}_{i\kp}^{-1}(t,\tp)}{\delta\eta_\jp(t)} \tilde{G}_{\kp j}(t,\tp).
\end{eqnarray}
Using (\ref{diff-1}) in (\ref{sigma}), and doing some algebra we obtain
\begin{eqnarray}
\label{sigma-f}
\tilde{\Sigma}_{ij}(t,\tp) &=& {\cal W}_{i\kp\ip_1\jp_1}(\tp_1 t,t)\tilde{G}_{\kp i_1}
{\cal L}_{i_1 j\ip_1\jp_1}(t_1,\tp_1,\tp_1)\nonumber\\
&-&i \eta_i^*(t)\langle c_\jp^\dag(\tp_1)\rangle_\eta \tilde{G}_{\jp j}^{-1}(\tp_1,\tp)+
{\cal A}_{ij}(t,\tp)\nonumber\\
\end{eqnarray}
where
\begin{eqnarray}
{\cal W}_{ijkl}(t_1,t_2,t_3)&=& \bar{V}_{\ip i\jp j}\frac{\delta^2v_{kl}(t_1)}{\delta\eta_\ip^*(t_3)\eta_\jp(t_3)}\label{n-fun-1}\\
{\cal L}_{ijkl}(t_1,t_2,t_3)&=& -\frac{\delta\tilde{G}_{ij}^{-1}(t_1,t_2)}{\delta v_{kl}(t_3)}\label{n-fun-2}
\end{eqnarray}
which for $\{\eta\}=0$ reduce to the screened Coulomb potential and the vertex function \cite{gunnarsson,hedin}, respectively, and
\begin{widetext}
\begin{eqnarray}
\label{aij}
{\cal A}_{ij}(t,\tp)&=&
i\bar{V}_{\ip i\jp\kp}\tilde{G}_{\kp j_1}(t,\tp_1)
\left(\frac{\delta\tilde{G}_{j_1j}^{-1}(\tp_1,\tp)}{\delta \eta_\jp(t)}\langle c_\jp^\dag(t)\rangle_\eta
+\langle c_\jp(t)\rangle_\eta\frac{\delta\tilde{G}_{j_1j}^{-1}(\tp_1,\tp)}{\delta \eta_\ip^*(t)}
+\right)\nonumber\\
&-& \bar{V}{\ip i\jp\kp}\tilde{G}_{\kp i_1}(t,t_1)\frac{\delta^2 \tilde{G}_{i_1j}^{-1}(t_1,\tp)}
{\delta v_{\ip_1\jp_1}(\tp_1)\delta v_{i_1j_1}(\tp_2)}\frac{\delta v_{\ip_1\jp_1}(\tp_1)}
{\delta\eta^*_\ip(t)}\frac{\delta v_{i_1j_1}(\tp_2)}{\delta\eta_\jp(t)}\nonumber\\
&+& \left(\bar{V}_{\ip i\jp\kp}\frac{\delta \tilde{G}_{\kp i_1}(t,t_1)}{\delta \eta^*_\ip(t)}
\frac{\delta \tilde{G}_{i_1j}^{-1}(t_1,\tp)}{\delta \eta_\jp(t)}
+\eta_\ip \Leftrightarrow \eta_\jp\right).
\end{eqnarray}
\end{widetext}
When $\{\eta\}=0$, ${\cal A}_{ij}$ vanishes and only the first term on the rhs of Eq. (\ref{sigma-f}) survives.

Using Eqs. (\ref{dyson-1}) and (\ref{hartree}) in (\ref{n-fun-2}) and (\ref{n-fun-1}) respectively 
and making use of the identity (\ref{identity}), we obtain a closed equation for ${\cal L}_{ijkl}$ and ${\cal W}_{ijkl}$
\begin{widetext}
\begin{eqnarray}
{\cal L}_{ijkl}(t_1,t_2,t_3)&=& -\delta_{ik}\delta_{lj}\delta(t_1-t_3)\delta(t_1-t_2)\label{lambda-final}\\
&+&\frac{\delta\tilde{\Sigma}_{ij}(t_1,t_2)}{\delta\tilde{G}_{\ip\jp}(\tp_1,\tp_2)} 
\tilde{G}_{\ip\jp_1}(\tp_1,\tp_3){\cal L}_{\jp_1\jp_2kl}(\tp_3,\tp_4,t_3)\tilde{G}_{\jp_2\jp}(\tp_4,\tp_2)\nonumber\\
{\cal W}_{ijkl}(t_1,t_2,t_3)&=& {\cal Q}_{ijkl}(t_1,t_2,t_3)\label{w-final}\\
&-& i\bar{V}_{\ip_1 k\jp_1l}{\cal W}_{ij\ip_1\jp_1}(\tp_3,t_2,t_3)\tilde{G}_{\jp_1i_1}(t,\tp){\cal L}_{i_1k_1\ip_1\jp_1}(\tp_1,\tp_2,\tp_3)\tilde{G}_{k_1\ip_1}(\tp_2,t_1)\nonumber
\end{eqnarray}
\end{widetext}
where
\begin{widetext}
\begin{eqnarray}
\label{fun-q}
{\cal Q}_{ijkl}(t_1,t_2,t_3) &=& i\bar{V}_{\ip_1 k\jp_1 l}\bar{V}_{\ip i\jp j} \tilde{G}_{\jp_1i_1}(t_1,\tp_1)
\left( \frac{\delta\tilde{G}_{i_1k_1}^{-1}(\tp_1,\tp_2)}{\delta\eta_\ip^*(t_2)}
+\frac{\delta \tilde{G}_{k_1\ip_1}(\tp_2,t_1)}{\delta\eta_\jp(t_3)} \right)\nonumber\\
&+& i\bar{V}_{\ip_1 k\jp_1 l}\bar{V}_{\ip i\jp j}
\tilde{G}_{\jp_1i_1}(t,\tp) \tilde{G}_{k_1\ip_1}(\tp_2,t_1)K_{i_1k_1\ip\jp}(\tp_1,\tp_2,t_3,t_3)\nonumber\\
&+&2i\bar{V}_{\ip i\jp j}\frac{\delta^2}{\delta\eta_\ip^*(t_2)\delta\eta_\jp(t_3)}
\left[ \eta_\ip^*(t_1)\left(\langle c_\ip^\dag(t_1)\rangle+\frac{\delta}{\delta\eta_\ip^*(t_1)}\right)\delta_{kl}+c.c.\right].
\end{eqnarray}
\end{widetext}

On taking the limit $\{\eta\}=0$, Eqs. (\ref{dyson-1}), (\ref{sigma-f}) reduce to a closed set of equations for the Green's functions and the self-energy
\begin{eqnarray}
G_{ij}^{(m)}(t,\tp) &=& G_{ij}^{0(m)}(t,\tp)\nonumber\\
&+& G_{i\jp}^{0(m)}(t,t_1)\Sigma^{(m)}_{\jp\kp}(t_1,t_2)
G_{\kp j}^{(m)}(t_2,t) \label{closed-eq-1}\\
\Sigma_{ij}^{m}(t,\tp) &=&
S^{(m)}_{i\kp\ip\jp}(t_1 t,t)G^{(m)}_{\kp l}(t_1,t_2) \Lambda^{(m)}_{l j\ip\jp}(t_2,t_1,\tp) \label{closed-eq-2}\nonumber\\
\end{eqnarray}
where $S$ and $\Lambda$ are the screened Coulomb interaction and the vertex function, respectively, obtained from the reduction of Eqs. (\ref{lambda-final}) and (\ref{w-final})
\begin{widetext}
\begin{eqnarray}
\Lambda^{(m)}_{ijkl}(t_1,t_2,t_3)&=& -\delta_{ik}\delta_{lj}\delta(t_1-t_3)\delta(t_1-t_2)\nonumber\\
&+&\frac{\delta\Sigma^{(m)}_{ij}(t_1,t_2)}{\delta G^{(m)}_{\ip\jp}(\tp_1,\tp_2)} G^{(m)}_{\ip\jp_1}(\tp_1,\tp_3) \Lambda^{(m)}_{\jp_1\jp_2kl}(\tp_3,\tp_4,t_3)G^{(m)}_{\jp_2\jp}(\tp_4,\tp_2)\label{closed-eq-3}\\
 S^{(m)}_{ijkl}(t_1,t_2,t_3)&=& -i\bar{V}_{\ip_1 k\jp_1l}S^{(m)}_{ij\ip_1\jp_1}(\tp_3,t_2,t_3)G^{(m)}_{\jp_1i_1}(t,\tp) \Lambda^{(m)}_{i_1k_1\ip_1\jp_1}(\tp_1,\tp_2,\tp_3)G^{(m)}_{k_1\ip_1}(\tp_2,t_1)\label{closed-eq-4}
\end{eqnarray}
\end{widetext}
The set of Eqs. (\ref{closed-eq-1})-(\ref{closed-eq-4}) is exact.

We next derive a closed equation for the two-particle Green's function in the presence of a core-hole created at core orbital $m$.
This can be readily generalized to the case of two core-holes.

Equation (\ref{dyson-1}) can be written as
\begin{eqnarray}
\label{dyson-2}
\tilde{G}_{ij}(t,\tp)&=&\tilde{G}_{ij}^{0}(t,\tp)+\tilde{G}_{i\jp}^{0}(t,\tp_1)\tilde{\Sigma}_{\jp\kp}(\tp_1,\tp_2)
\tilde{G}_{\kp j}(\tp_2,\tp).\nonumber\\
\end{eqnarray}
Differentiating Eq. (\ref{dyson-2}), first with respect to $\eta_k(t_3)$ and then with respect to $\eta^*_l(t_4)$, taking the limit $\{\eta\}=0$ and making use of Eq. (\ref{111}), we get
\begin{eqnarray}
\label{2-part-1}
&&[\delta_{i\kp}\delta(t_1-\tp_2)-G_{i\jp}^{0(m)}(t_1,\tp_1)\Sigma^{(m)}_{\jp\kp}(\tp_1,\tp_2)]
\nonumber\\
&\times& \cg_{\kp jkl}^{(m)}(\tp_2,t_2,t_3,t_4)=iG_{il}^{0(m)}(t_1,t_4)G^{(m)}_{kj}(t_3,t_2)
\nonumber\\
&+& G_{i\jp}^{0(m)}(t_1,\tp_1) G_{\kp j}^{(m)}(\tp_2,t_2)\Xi_{\jp\kp lk}^{(m)}(\tp_1,\tp_2,t_3,t_4)\nonumber\\
\end{eqnarray}
where
\begin{eqnarray}
\label{xi}
&&\Xi_{ijkl}^{(m)}(t_1,t_2,t_3,t_4)= \left.\frac{\delta^2 \tilde{\Sigma}_{ij}(t_1,t_2)}
{\delta\eta_k^*(t_3)\delta\eta_l(t_4)}\right|_{\{\eta\}=0}\nonumber\\
&=& \left[ \frac{\delta}{\delta\eta_k^*(t_3)}\left(\frac{\delta\tilde{\Sigma}_{ij}(t_1,t_2)}
{\delta \tilde{G}_{\kp\lp}(\tp_1,\tp_2)}\frac{\delta \tilde{G}_{\kp\lp}(\tp_1,\tp_2)}{\delta\eta_l(t_4)}\right)\right]_{\{\eta\}=0}\nonumber\\
&=& \frac{\delta\Sigma_{ij}(t_1,t_2)}{\delta G_{\kp\lp}(\tp_1,\tp_2)}\cg_{\kp\lp kl}^{(m)}(\tp_1,\tp_2,t_3,t_4)
\end{eqnarray}
where in going from second to the third line we used the fact that,
$(\delta\tilde{G}_{ij}/\delta\eta_k)_{\{\eta\}=0}=0$.

Substituting Eq. (\ref{xi}) in (\ref{2-part-1}) and using Eq. (\ref{closed-eq-1}) to replace the first line in Eq. (\ref{2-part-1}), we get
\begin{widetext}
\begin{eqnarray}
\label{2-part-2}
\cg_{ijkl}^{(m)}(t_1,t_2,t_3,t_4)&=& iG_{kj}^{(m)}(t_3,t_2)G_{il}^{(m)}(t_1,t_4)\nonumber\\
&+& G_{i\ip}^{(m)}(t_1,\tp_1)G_{\kp j}^{(m)}(\tp_2,t_2)
\Xi^{(m)}_{\ip\kp\ip_1\kp_1}(\tp_1,\tp_2,\tp_3,\tp_4)
\cg_{ijkl}^{(m)}(t_1,t_2,t_3,t_4)
\end{eqnarray}
\end{widetext}
where kernel,
\begin{eqnarray}
 \label{kernel-1}
\Xi^{(m)}_{\ip\kp\ip_1\kp_1}(\tp_1,\tp_2,\tp_3,\tp_4)= \frac{\delta\Sigma^{(m)}_{\ip\kp}(\tp_1,\tp_2)}{\delta G_{\lp_1\kp_1}^{(m)}(\tp_3,\tp_4)}
\end{eqnarray}

Equations (\ref{closed-eq-1})-(\ref{closed-eq-4}) together with Eq. (\ref{2-part-2}) and (\ref{kernel-1}) constitute the set of equations
(\ref{21aa}) - (\ref{2-electron}) used in the main text.

%\begin{thebibliography}{99}

\end{document}